# Comment on "Mean First Passage Time for Nuclear Fission and the Emission of Light Particles"

**Abstract:** Theoretical methods, interpretations and conclusions on the fission dynamics in a recent publication of H. Hofmann and F. A. Ivanyuk on the mean first passage time are critically considered.

*Motivation.* H. Hofmann and F. A. Ivanyuk [1] published a critical assessment of the observability of transient effects by measuring particle evaporation in fission. They claimed that the problem is badly defined due to remarkable uncertainties in the initial conditions and considerable arbitrariness in choosing time zero. Moreover, it was claimed that there is no room for transient effects and for saddle-to-scission times in the concept of the mean first passage time (MFPT) and that it is not allowed to consider relaxation effects in the fission decay using statistical codes. We give a critical comment on these statements.

*Initial conditions*. Previous studies on dissipation in fission were performed with either one of two equivalent approaches [2], the Langevin equation, allowing for evaporation along the dynamic trajectory in deformation space, or by incorporating a time-dependent fission-decay width from the Fokker-Planck equation in an evaporation calculation. They clearly demonstrated relaxation effects in the evolution of the fission process. The initial properties of the nuclear system are defined by the entrance channel. They are not "at one's disposal", as claimed in ref. [1].

*Mean first passage time.* Ref. [1] cites some properties of the MFPT, which are only valid in the case of small noise ($T << B_f$) [3], and claims that there is no room for transient effects and dynamic saddle-to scission times in general. The absorption at the exit point in the definition of the MFPT is taken as an argument in [1] that "it is not permitted to use in (3) the currents shown in Fig. 1" to calculate the MFPT. We argue that this is not true for the exit point at scission, where trajectories with negative velocity are rare. Here, the closed expression, relation (2) of ref. [1], is the solution of equation (3) for the over-damped case, using the currents given by the time-dependent solution of the Smoluchowski equation, which clearly show transient effects and saddle-to-scission times as demonstrated in figure 1. Thus, they are also present in the MFPT expressed by equation (2). Absorption at smaller deformations, where negative velocity components are not negligible, does not remove relaxation effects but leads to an unphysical change of the properties of the system due to the additional "leakage" introduced at the absorption point. Therefore, one should be careful in studying the dynamics of the system by means of the MFPT. Also the restriction to one starting point in the definition of the MFPT and the lacking consideration of evaporation hamper realistic studies of the fission dynamics using the MFPT.

The relations MFPT($Q_a -> Q_b$) = ½$\tau_k$ and MFPT($Q_a -> Q_s$) = $\tau_k$ have been derived for the case of small noise [3]. They are *not* generally valid. The saturation of the MFPT for large deformations (figure 3 in ref. [1]) is a trivial consequence of the strongly increasing negative slope of the cubic potential for deformations beyond the saddle point. For a parabolic potential, this feature does not appear. The asymptotic saturation of the MFPT at large deformations for a cubic or higher-order potential is not an argument against the existence of a dynamic saddle-to-scission time.

*Correlation time.* It is claimed that the distributions of starting points at $t_1$ for Langevin trajectories passing by the saddle at $t_2$ for two different time intervals $\Delta_t = \{t_1 \to t_2\}$ in figure 2 of ref [1] prove "considerable arbitrariness in choosing time zero" for calculating transient effects. We argue that this statement confounds the correlation between the starting point at $t_1$ and the point reached at $t_2$ of individual trajectories during a characteristic time interval $\Delta_t$, which is demonstrated by figure 2 and which does not depend on the starting time $t_1$, from transient effects, which are a relaxation phenomenon of the probability distribution of the whole statistical ensemble towards thermal quasi-equilibrium.

In addition, the formulation of ref. [1], that the MFPT "takes into account an average of all initial points". is not justified by Fig. 2. Indeed, by definition of the MFPT all trajectories start exactly at the same initial point $Q_a$ [3]. The MFPT is in general sensitive to the value of $Q_a$; only in the case of small noise this sensitivity is weak.

***Conclusions.*** The MFPT is clearly influenced by relaxation effects. The MFPT characterizes the dynamic evolution of the fissioning system under the influence of dissipation only, if evaluated at or close to the scission point. For smaller deformation values, the assumption of absorption at the deformation of interest is in severe contradiction to the physics of the problem. In these cases, the decay rate should be considered. The closed expression for the MFPT is of limited use due to the several severe approximations applied in its derivation. In addition, it only gives a mean value and thus cannot describe the detailed features of the time dependence of passage times, which are important for an application to a realistic case including particle evaporation.

K.-H. Schmidt[1], J. Benlliure[2], D. Boilley[3], A. Heinz[4], A. Junghans[5], B. Jurado[3], A. Kelic[1], J. Pereira[2], C. Schmitt[1], O. Yordanov[1]
[1]*GSI Planckstr. 1, D- 64291 Darmstadt,* [2]*Univ. de Santiago de Compostela, E-15706,* [3]*GANIL, BP 55027, F-14076 Caen cedex 05,* [4]*Yale University, New Haven, CT 06520,* [5]*FZ Rossendorf, PF 510119, D-01314 Dresden*
PACS numbers: 24.75.+i, 05.60.-k, 24.10.pa, 24.60.Dr

[1] H. Hofmann, F. A. Ivanyuk, Phys. Rev. Lett. 90 (2003) 132701.
[2] H. A. Weidenmüller, Prog. Part. Nucl. Phys., Pergamon, Oxford, Vol. 3 (1980) 49.
[3] P. Hänggi, P. Talkner, B. Morkovec, Rev. Mod. Phys. 62 (1990) 251.